\documentclass[11pt,twoside]{article}

\usepackage{asp2006}
\usepackage{epsf}
\usepackage{psfig}
\usepackage{lscape}

\begin{document}
\title{An on-going multi-wavelength survey of Ly$\alpha$ emitters at redshifts $z = 2 - 8$}   
\author{Kim K. Nilsson,$^{1,2}$ Johan P.U. Fynbo$^2$, Palle M{\o}ller$^1$ and Alvaro Orsi$^3$}  
\affil{$^1$ESO Garching, Karl-Scwarzschild-Strasse 2, D-85748 Garching bei M{\"u}nchen, Germany} 
\affil{$^2$Dark Cosmology Centre, Juliane Maries Vej 30, DK-2100 K{\o}benhavn {\O}, Denmark}
\affil{$^3$Department of Astronomy, Pontificia Universidad Cat{\'o}lica, Casilla 306, Santiago 22, Chile}

\begin{abstract}
In the last decade, the technique of finding Ly$\alpha$-emitters 
through narrow-band imaging has become a promising method of detecting
high redshift galaxies. Ly$\alpha$-emitters have been found from redshifts
$z \ga 2$, up to the highest redshift source known to date at $z = 6.96$. 
Several surveys are also underway to find $z = 7 - 9$ sources. 
But these very high redshift sources are too faint to be studied in great
detail, and more information can be found from studying the same class of
objects at lower redshifts. Here we present our survey strategy to determine
the nature of Ly$\alpha$-emitters at lower redshifts, through multi-wavelength
surveys, and our plans to extend the survey to redshift $z = 8.8$. 
\end{abstract}

\section{Introduction}
The success of narrow-band searches, focusing on Ly$\alpha$, was
predicted by Partridge \& Peebles (1967) nearly 40 years ago, but early 
surveys failed to produce anything other than
upper limits. The unexpected faintness of the objects caused it to
take almost three decades before the narrow-band
technique was successfully used to identify the first
high redshift Ly$\alpha$ emitting galaxies (M{\o}ller \& Warren 1993; 
Hu \& McMahon 1996), and
it is only recently, with the advent of 8 m class telescopes and sensitive
detectors,
that larger samples of Ly$\alpha$ selected objects have been reported
(Steidel et al. 2000; Malhotra \& Rhoads 2002; Fynbo et al. 2003; Venemans et 
al. 2005). Lately, much of the attention
has been turned to detecting Ly$\alpha$-emitters in the redshift range of
$z = 4 - 7$, and the lower redshifts have been neglected.
Due to the increase in luminosity distance, objects at these redshifts are
very difficult to study in greater detail, once they are found. At lower
redshifts (i.e. $z = 2 - 3$), it is still possible to detect, or at least put
strong upper limits on, the SED of the galaxy. Hence at these redshifts, we can 
study the nature of Ly$\alpha$ selected galaxies, using public 
multi-wavelength surveys. However,
Ly$\alpha$ also appears to be an excellent tool in finding very high redshift
sources (e.g. Iye et al. 2006) and so it is also of interest to push
the limits of what is observable. We can then use the strength
of wide-field narrow-band imaging for Ly$\alpha$ to find very high redshift
sources ($z > 8$), which in turn will enable a study of when re-ionisation
happened and how galaxies formed.

In these proceedings, we present our on-going survey strategy to study the 
nature of Ly$\alpha$-emitters at lower redshifts, and present a
future Public Survey that we expect will find several very high redshift sources.
Magnitudes are in the AB system.

\subsection*{Survey plans}
In order to understand the physical nature of any high redshift galaxy,
multi-wavelength coverage enabling SED fitting analysis is crucial. Hence
recent large surveys, such as the GOODS fields (Giavalisco et al. 2004) and 
the COSMOS\footnote{http://www.astro.caltech.edu/\~~cosmos/} survey are 
invaluable. These are the fields we target in our 
narrow-band surveys. We have nearly completed a survey at redshift $z = 3.15$
in the GOODS-S field (Nilsson et al. 2006; Nilsson et al. 2007, to be 
submitted). The narrow-band imaging we have in this field
comes from the VLT/FORS instrument, with a field-of-view of $7'\times7'$. In this 
field we have found 23 candidate compact Ly$\alpha$-emitting candidates,
and one so-called Ly$\alpha$-blob. We have also started a similar survey in 
the GOODS-N field, using the Nordic Optical Telescope (NOT) on La Palma, 
Canary Islands, Spain. This survey is focused on two redshifts, 
$z = 2.34$~and~3.15. The data shows 10 candidates in the lower redshift filter. 
The higher redshift
filter imaging is not complete. Finally, we are proposing a Public Survey with
the VISTA telescope; ELVIS - Emission Line galaxies with VISTA. This 
survey will be conducted with a narrow-band filter centred on Ly$\alpha$ at 
$z = 8.8$, see section~\ref{nil:sec3}

\section{GOODS-S and the Ly$\alpha$-blob}\label{nil:sec2}
In December 2002, one $7'\times7'$ field, centred on RA~$03^\mathrm{h}32^\mathrm{m}21.8^\mathrm{s}$ and Dec~$-27^\circ 45' 52''$ (J2000) was observed with
VLT/FORS1. The field was observed for 8.3 hours in a narrow-band filter
with central wavelength 5055~{\AA} and with a FWHM of 60~{\AA}. This 
corresponds to a redshift range of $z = 3.126 - 3.174$ for Ly$\alpha$. 
The images were reduced and calibrated with standard methods (see Nilsson et
al. 2007, to be submitted, for details). The 5$\sigma$ limiting magnitude
for the narrow-band image is 25.8.

We selected objects using the HST/ACS F435W and F606W  
images available from GOODS and using SExtractor. The narrow-band image
was used as a detection image, and broad-band magnitudes were measured in
circular apertures placed at the
position of the narrow-band sources on rebinned HST images. Objects with a 
measured equivalent width larger
than zero and ``signal-to-noise'' of the equivalent width measurement larger
than three were selected. Our final catalog includes 26 objects, of which three 
are associated with the 
Ly$\alpha$-blob (sec.~\ref{nil:sec2:1}; Nilsson et al. 2006) and the other 23 are
compact Ly$\alpha$-emitting candidates which will be presented in a forthcoming
paper (Nilsson et al. 2007, to be submitted).

\subsection{The Ly$\alpha$ blob}\label{nil:sec2:1}
\begin{figure}[!ht]
\plotone{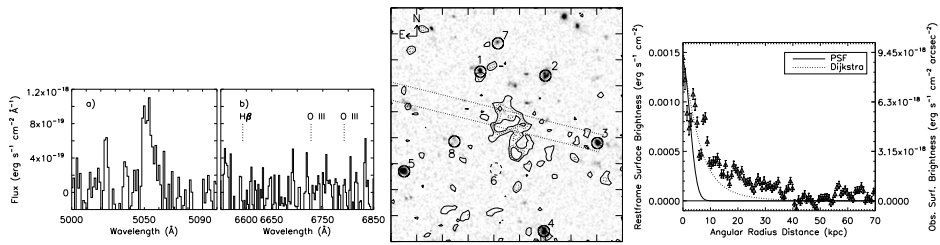}
\caption{\emph{Left}~\emph{a)} Flux calibrated spectrum of the blob emission
line. The
line has the characteristic blue side absorption, indicating high redshift.
\emph{b)} The part of the spectrum where H$\beta$ and [OIII] should have been 
observed if the emission line was [OII] at a redshift of $z \approx 0.36$.
These lines are not observed and
therefore we conclude the observed line is due to Ly$\alpha$ at $z=3.16$.
\emph{Middle} Contour-plot of narrow-band emission from the Ly$\alpha$ blob
overlaid the
HST V-band image. The narrow-band image has been continuum
subtracted by subtracting the re-binned, smoothed and scaled HST/V-band image.
Contour levels are $2 \cdot 10^{-4}$, $4 \cdot 10^{-4}$ and $6
\cdot 10^{-4}$~erg~s$^{-1}$~cm$^{-2}$ in restframe flux (corresponding
to $1.2 \cdot 10^{-18}$, $2.5 \cdot 10^{-18}$ and $3.7 \cdot 10^{-18}$ in
observed flux). The image is $18'' \times 18''$ ($18''$ corresponds to a 
physical size of $\sim 133$~kpc at $z = 3.15$). The dotted lines indicate
the slitlet position for our follow-up spectroscopy.
\emph{Right} Plot of surface brightness as function of radius.
The flux is the sky subtracted narrow-band flux.
The PSF of the image is illustrated by the solid line, and the
dotted line is the best fit model of Dijkstra et al. (2006a,b).
The deficit at $\sim 45$~kpc is due to the asymmetric appearance of
the blob.}\label{nil:fig1}
\end{figure}

\begin{figure}[!ht]
\plotone{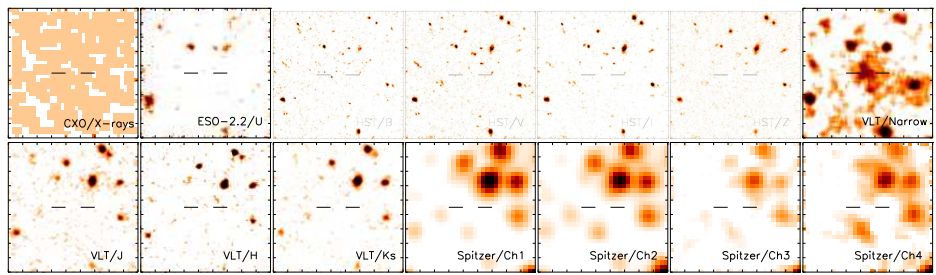}
\caption{Thumbnail images of all available multi-wavelength data in the GOODS 
South field, centred on the Ly$\alpha$ blob. All images are $18'' \times 18''$.}\label{nil:fig2}
\end{figure}

In the narrow-band image, we found a so-called Ly$\alpha$-blob, similar to
what has been found by many authors before (e.g. Steidel et al. 2000; Matsuda 
et al. 2004). The 
blob has a size
of approximately 70~kpc in projected diameter and a total Ly$\alpha$ line
luminosity of $1.0 \cdot 10^{43}$~erg~s$^{-1}$. A contour-plot, as well as a
plot of the blob surface brightness and the spectrum can be found in 
figure~\ref{nil:fig1}. The total multi-wavelength data on the blob can be seen 
in figure~\ref{nil:fig2}.

There have been three proposed energy sources of blobs. These are \emph{i)} 
hidden QSOs (Haiman \& Rees 2001; Weidinger et al. 2005), \emph{ii)} star 
formation and superwinds from
(possibly obscured) starburst galaxies (Taniguchi et al. 2001; Ohyama et al. 
2003), and \emph{iii)} 
so-called cold accretion 
(Haiman, Spaans \& Quataert 2000; Fardal et al. 2001; Sommer-Larsen 2005;
Dijkstra et al. 2006a,b; Dekel \& Birnboim 2006).
Cooling flows are phenomena observed in galaxy clusters for more than a decade.
These are explained by gas which is cooling
 much faster than the Hubble time through X-ray emission in the centres of the 
clusters. However, cooling emission from a galaxy, or a group sized halo can be 
dominated by Ly$\alpha$ emission, so-called cold accretion.
Until the discovery of our GOODS-S blob, cold accretion had not been observed.
Because of the extensive multi-wavelength coverage of our blob, we were
able to conclude that it is powered by cold accretion. 

The argument that the GOODS-S blob is powered by cold accretion comes from
ruling out the other two options. As seen in the thumb-nail images in 
figure~\ref{nil:fig2}, there are no detected continuum counterparts to the
blob Ly$\alpha$ emission. A starburst galaxy would have been observed in the
deep optical or IR data we have, and an AGN would have been detected in the
X-rays or the IR. Even an obscured AGN should be observed with the 
\emph{Spitzer/MIPS} bands. There were two objects of particular interest to us;
object~\#3 and object~\#6 in the contour-plot of figure~\ref{nil:fig1}. 
Object~\#3 since it appeared to have some faint Ly$\alpha$-emission surrounding
it, and had a photometric redshift similar to the blob 
($z_{phot} = 2.9^{+1.41}_{-0.59}$), and
object~\#6 since it only appeared in the \emph{Spitzer} bands. Object~\#3 does 
appear to have some narrow-band flux with very low significance
($< 2\sigma$) and so we studied the UV properties of this galaxy. However, the
measured UV flux of this galaxy is not sufficient to photo-ionise the blob. We
considered that object~\#6 was a hidden AGN, but its IR colours are more
indicative of a higher redshift starburst galaxy.
After ruling out the possibility that the blob was powered by star formation,
or AGN activity, we hence concluded that this blob must be the first observation
of cold accretion of neutral gas onto a dark matter halo.
For a more extensive discussion of this blob, see Nilsson et al. (2006). 

\section{ELVIS - Emission Line galaxies with VISTA Survey}\label{nil:sec3}
Ly$\alpha$-emitters is one of the most promising methods of finding very high
redshift ($z > 6$) galaxies. To date, the galaxy with highest redshift known
($z = 6.96$, Iye et al. 2006) was found as a Ly$\alpha$-emitter through 
narrow-band imaging techniques. Hence, there are now many proposed projects to 
find
$z > 7$ Ly$\alpha$-emitters. One such project is a Public Survey with the VISTA
telescope. VISTA\footnote{www.vista.ac.uk} is a new telescope being built near 
Paranal in Chile. It is a 
4-m telescope, with a wide-field IR camera array. It is equipped with the 
standard set of broad-band filters, as well as a custom-made narrow-band filter
with central wavelength 1185~nm and FWHM~$\approx 10$~nm
purchased by the Dark Cosmology Centre (DK). VISTA will devote 85~\% of its time
for Public Surveys and will be run by ESO. The Public Survey we propose is called 
ELVIS -- Emission Line galaxies with VISTA Survey (Nilsson et al. 2007, in 
prep.), now part of a larger survey called Ultra-VISTA, which proposes to do 
both very deep broad- and narrow-band imaging.

\begin{figure}[!ht]
\plottwo{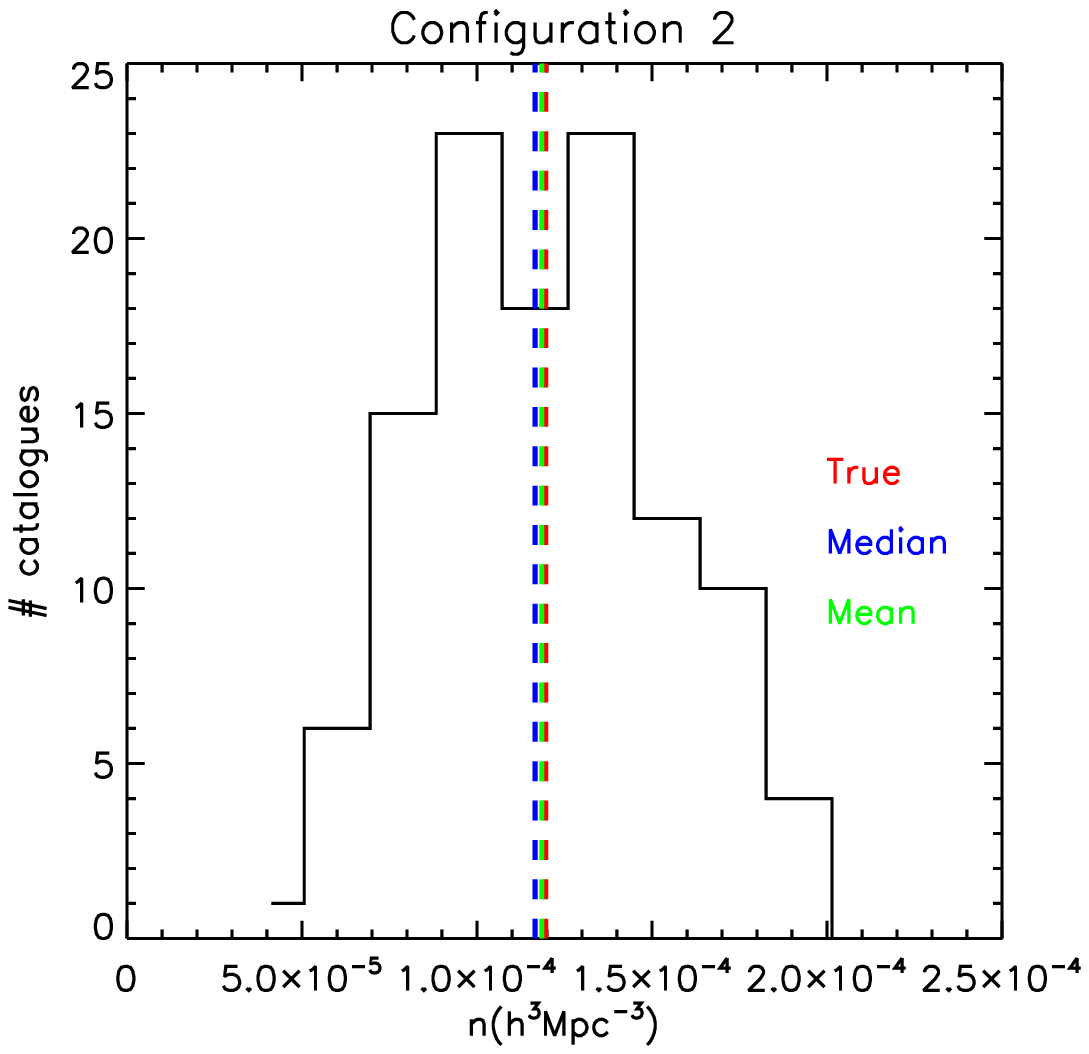}{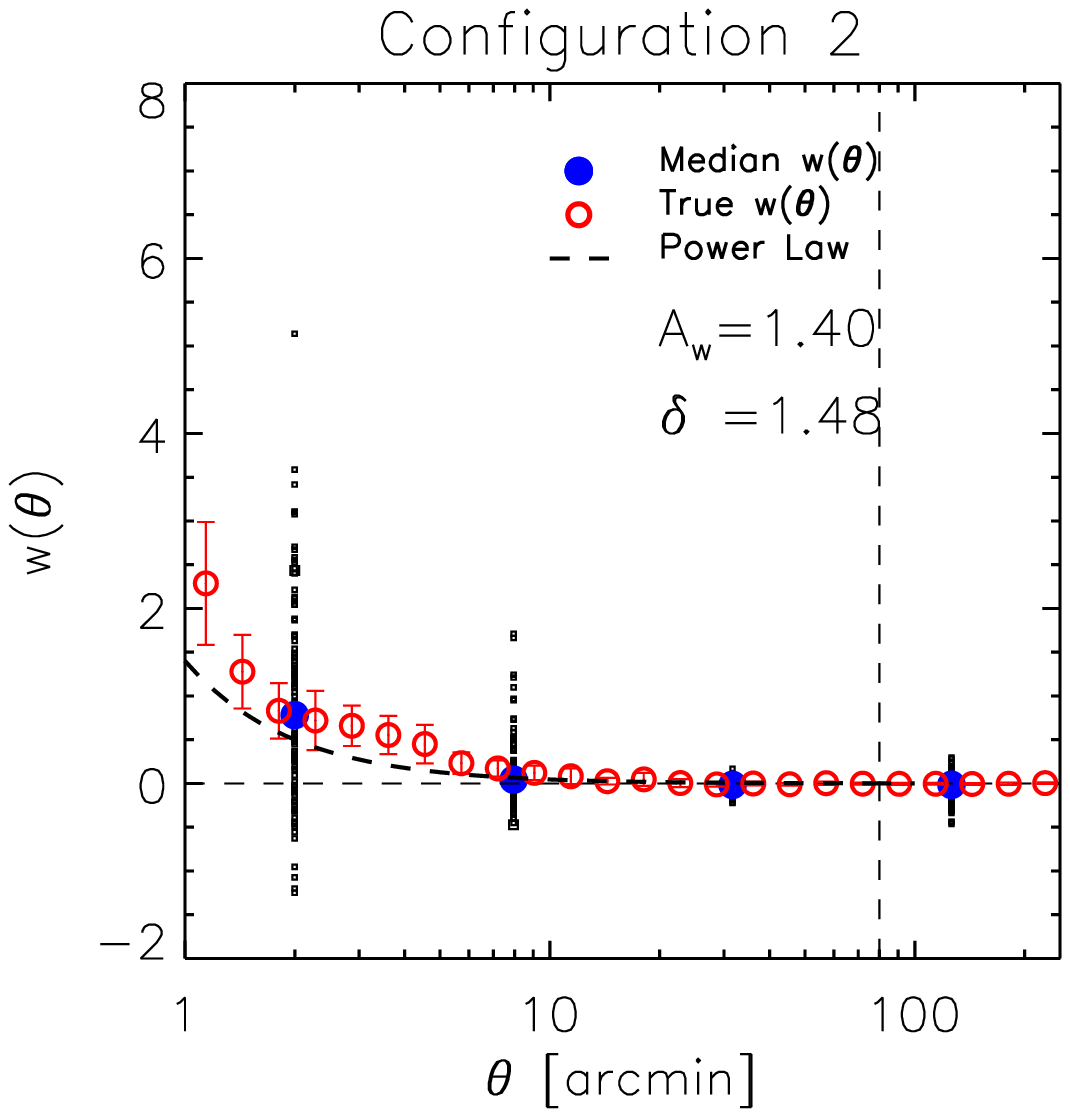}
\caption{\emph{Left} Number density of Ly$\alpha$-emitters at redshift $z = 8.8$
to the expected flux limit of ELVIS and an EW~$> 30$~{\AA}. We have made 112 mock
catalogs of Ly$\alpha$-emitters at $z = 8.8$ by incorporating the Millennium 
simulation to the GALFORM semi-analytical model and
extracted the number of objects detected in each catalog. This plot shows the
histogram of that number. The survey volume of
ELVIS is 137500~Mpc$^3$, hence we expect to detect approximately 15 objects
in the whole survey. \emph{Right} Angular correlation function. Blue solid dots
correspond to the median of the ``measured'' correlation function, red open 
circles correspond to the ``true'' correlation function, i.e. the whole catalog.
Plots from Nilsson et al. (2007) in prep. }\label{nil:fig3}
\end{figure}

The ELVIS part of the proposal, the deep narrow-band imaging, is planned to go 
to a depth of $3.7 \cdot 10^{-18}$~erg~cm$^{-2}$~s$^{-1}$ (5$\sigma$) in an
0.9~square degree field. The science case is to find a large sample of 
emission-line galaxies at several redshifts. The observations will be sensitive
to H$\alpha$ at redshift $z = 0.8$, [OIII] and H$\beta$ at $z = 1.4$, [OII] at
$z = 2.2$ and Ly$\alpha$ at redshift $z = 8.8$. For the Ly$\alpha$-emitters,
a prediction of the number counts of objects found can be seen in 
figure~\ref{nil:fig3}. We expect to detect approximately 15 such objects in our 
survey.  

These Ly$\alpha$-emitters can then be used to study the re-ionisation state
of the Universe at this redshift, and to study clustering, as seen
in figure~\ref{nil:fig3}. All of the emission lines surveyed with ELVIS are
also good star formation rate indicators. Thus, the survey will enable a 
detailed study of the star formation history of the Universe. For H$\alpha$, 
the survey will reach more than three orders of magnitude below L* and so ELVIS
will probe the very faint end of the H$\alpha$ luminosity function at $z = 0.8$.

\section{Summary}
To summarise, we have an extensive survey plan underway to determine the
nature of Ly$\alpha$-emitters at redshifts $z \sim 2 - 3$, using public,
multi-wavelength data in surveys such as GOODS and COSMOS. We do this by e.g.
SED fitting analysis and morphological analysis from HST data. In our 
GOODS-S survey, we discovered a so-called Ly$\alpha$-blob. Its properties
are best explained by cold accretion onto a dark matter halo. This in turn 
means that the energy sources of these blobs are diverse, since previous 
authors have found that other blobs have been better explained by star 
formation (Steidel et al. 2000; Francis et al. 2001; Matsuda et al. 2004; 
Palunas et al. 2004) or AGN activity 
(Fynbo et al. 1999; Keel et al. 1999; Dey et al. 2005; Villar-Martin et al.
2005; Weidinger et al. 2005). Larger samples of this rare 
phenomenon are
required before we can fully understand the underlying processes in these
objects. Finally, we have also presented our future Public Survey (ELVIS),
planned to start observing in the winter of 2008. It is expected to
discover tens of Ly$\alpha$-emitters at $z = 8.8$, as well as hundreds of
intermediate redshift emission-line galaxies and thousands of 
H$\alpha$-emitters at lower redshift. 

\acknowledgements 
K.N. gratefully acknowledges support from IDA - Instrumentcentre for 
Danish Astrophysics. The Dark Cosmology Centre is funded by the DNRF.


\begin{thebibliography}{}
\bibitem[Dekel \& Birnboim (2006)]{dekel06}
Dekel, A. \& Birnboim, Y., 2006, MNRAS, 368, 2
\bibitem[Dey et al. (2005)]{dey05}
Dey, A., Bian, C., Soifer, B.T. et al. 2005, ApJ, 629, 654
\bibitem[Dijkstra et al. (2006a)]{dijkstra06a}
Dijkstra, M., Haiman, Z. \& Spaans, M. 2006, ApJ, 649, 14
\bibitem[Dijkstra et al. (2006b)]{dijkstra06b}
Dijkstra, M., Haiman, Z. \& Spaans, M. 2006, ApJ, 649, 37
\bibitem[Fardal et al. (2001)]{fardal01}
Fardal, M.A., Katz, N., Gardner, J.P. et al., 2001, ApJ, 562, 605
\bibitem[Francis et al. (2001)]{francis01}
Francis, P.J., Williger, G.M. \& Collins, N.R. 2001, ApJ, 554, 1001
\bibitem[Fynbo et al. (1999)]{fynbo99}
Fynbo, J.P.U., M{\o}ller, P. \& Warren, S.J. 1999, MNRAS, 305, 849
\bibitem[Fynbo et al. (2003)]{fynbo03}
Fynbo, J.P.U., Ledoux, C., M{\o}ller, P., Thomsen, B. \& Burud, I. 2003, A\&A, 407, 147
\bibitem[Giavalisco et al.~(2004)]{gia04}
Giavalisco, M., Ferguson, H. C., Koekemoer, A. M. et al. 2004, ApJ, 600, L93
\bibitem[Haiman, Spaans \& Quataert (2000)]{haiman00}
Haiman, Z., Spaans, M. \& Quataert, E., 2000, ApJL 537, L5
\bibitem[Haiman \& Rees (2001)]{haiman01}
Haiman, Z. \& Rees, M.J., 2001, ApJ 556, 87
\bibitem[Hu \& McMahon (1996)]{hu96}
Hu, E.M. \& McMahon, R.G. 1996, Nature, 382, 231
\bibitem[Iye et al. (2006)]{iye06}
Iye, M., Ota, K., Kashikawa, N. et al. 2006, Nature, 443, 186
\bibitem[Keel et al. (1999)]{keel99}
Keel, W.C., Cohen, S.H., Windhorst, R.A. \& Waddington, I. 1999, AJ, 118, 2547
\bibitem[Malhotra \& Rhoads (2002)]{malhotra02}
Malhotra, S., \& Rhoads, J.E. 2002, ApJL, 565, L71
\bibitem[Matsuda et al. (2004)]{matsuda04}
Matsuda, Y., Yamada, T., Hayashino, T. et al. 2004, AJ, 128, 569 
\bibitem[M{\o}ller \& Warren (1993)]{moller93}
M{\o}ller, P. and Warren, S.J. 1993, A\&A, 270, 43
\bibitem[Nilsson et al. (2006)]{nilsson06}
Nilsson, K.K., Fynbo, J.P.U., M{\o}ller, P., Sommer-Larsen, J., \& Ledoux, C. 2006, A\&A, 452, L23
\bibitem[Ohyama et al. (2003)]{ohyama03}
Ohyama, Y., Taniguchi, Y., Kawabata, K.S. et al., 2003, ApJ, 591, L9
\bibitem[Palunas et al. (2004)]{palunas04}
Palunas, P., Teplitz, H.I., Francis, P.J., Williger, G.M \& Woodgate, B.E. 2004, ApJ, 602, 545 
\bibitem[Partridge \& Peebles (1967)]{part67}
Partridge, R.B. \& Peebles, P.J.E. 1967, ApJ, 147, 868
\bibitem[Sommer-Larsen (2005)]{somm05}
Sommer-Larsen, J., 2005, in the proceedings of the ``Island Universes: Structure and Evolution of Disk Galaxies'' conference held in Terschelling, Netherlands, July 2005, ed. R de Jong (Springer Dordrecht), astro-ph/0512485
\bibitem[Steidel et al. (2000)]{steidel00}
Steidel, C.C., Adelberger, K., Shapley, A.E. et al. 2000, ApJ, 532, 170
\bibitem[Taniguchi et al. (2001)]{tani01}
Taniguchi, Y., Shioya, Y. \& Kakazu, Y., 2001, ApJL 562, L15
\bibitem[Venemans et al. (2005)]{vene05}
Venemans, B.P., R{\"o}ttgering, H.J.A., Miley, G.K. et al. 2005, A\&A, 431, 793
\bibitem[Villar-Martin et al. (2005)]{villmart05}
Villar-Martin, M., Sanchez, S.F. \& de Breuck, C. 2005, MNRAS, 359, L5
\bibitem[Weidinger et al. (2005)]{weid05}
Weidinger, M., M{\o}ller, P., Fynbo, J.P.U. \& Thomsen, B. 2005, A\&A, 436, 825
\end{thebibliography}
\end{document}